\documentclass[10pt,conference]{IEEEtran}
\IEEEoverridecommandlockouts

\usepackage{cite}
\usepackage{amsmath,amssymb,amsfonts}
\usepackage{graphicx}
\usepackage{booktabs}
\usepackage{url}
\usepackage{xcolor}
\usepackage{listings}
\usepackage{tikz}
\usetikzlibrary{arrows.meta,positioning,calc,fit}
\usepackage[hidelinks]{hyperref}

\newcommand{\tool}{OpenRE-Bench}

\newcommand{\auditPoolN}{1{,}396}
\newcommand{\auditPoolNOne}{958}
\newcommand{\auditPoolNTwo}{438}
\newcommand{\auditRetainedN}{50}
\newcommand{\auditRejectedN}{30}
\newcommand{\auditPrec}{84.0\%}
\newcommand{\auditPrecCI}{[71.5, 91.7]\%}
\newcommand{\auditFOR}{90.0\%}
\newcommand{\auditFORCI}{[74.4, 96.5]\%}

\newcommand{\auditPrecConsensus}{82.6\%}

\newcommand{\auditFORConsensus}{92.3\%}

\newcommand{\auditPoolPrec}{85.9\%}
\newcommand{\auditPoolPrecCI}{[78.1, 93.6]\%}
\newcommand{\auditDisagree}{8}
\newcommand{\auditKappa}{0.65}

\begin{document}

\title{\tool: A Benchmark Harness for\\
Multi-Agent Requirements Engineering\\
with Mined GitHub Issues}

\author{\IEEEauthorblockN{Haowei Cheng\IEEEauthorrefmark{1},
Milhan Kim\IEEEauthorrefmark{1},
Bowen Jiang\IEEEauthorrefmark{2},
Weixing Zhang\IEEEauthorrefmark{2},
Yuhong Fu\IEEEauthorrefmark{3},\\
Zhixu Liu\IEEEauthorrefmark{1},
Naoyasu Ubayashi\IEEEauthorrefmark{1},
Hironori Washizaki\IEEEauthorrefmark{1}}
\IEEEauthorblockA{\IEEEauthorrefmark{1}Waseda University, Tokyo, Japan\\
\IEEEauthorrefmark{2}Karlsruhe Institute of Technology (KIT), Germany\\
\IEEEauthorrefmark{3}Adelaide University, Adelaide, Australia}
}

\maketitle

\begin{abstract}
\tool{} is an open-source benchmarking infrastructure for large language model (LLM)
based multi-agent requirements-engineering (RE) frameworks. Such frameworks are
increasingly common, but they use different agent roles, negotiation policies, and
output shapes, and are typically evaluated on their own runtimes, metrics, and small
sets of curated system descriptions, which makes fair, reproducible comparison
difficult. \tool{} addresses
this with four reusable pieces: a shared run harness with one input, phase-artifact,
and metric contract; an adapter mechanism through which a framework joins by
implementing a single adapter, leaving cases, metrics, and other adapters untouched; a
reproducible GitHub issue-mining pipeline that grows the corpus from real feature
requests; and a versioned case suite. As a demonstration, we mine 3{,}000 fetched
issues across five projects (three languages, five domains) into five cases, package
them with five hand-built cases as the compact v1.0 suite, and run three heterogeneous
frameworks (MARE and iReDev as paper-faithful reimplementations, plus QUARE) under one
protocol. \tool{} originated as the evaluation back-end of our QUARE study and has
since been reused by two further studies of our group. It is released with a pinned
environment, a Docker image, frozen API snapshots, a 651-test suite, and the full pool
of \auditPoolN{} filtered issues behind the mined cases.
\end{abstract}


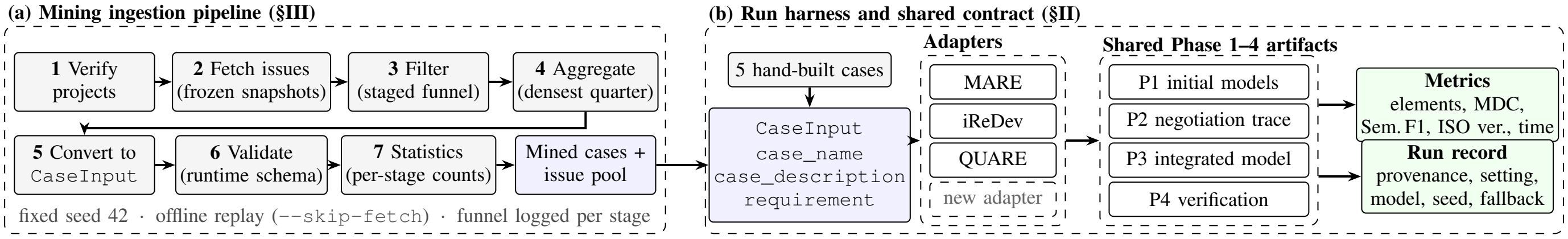
\begin{figure*}[t]
\centering
\resizebox{\textwidth}{!}{%
\begin{tikzpicture}[
  font=\scriptsize, >={Stealth[length=1.6mm]},
  stp/.style={draw, rounded corners=1.5pt, fill=gray!8, align=center,
    minimum width=1.72cm, minimum height=0.72cm, inner sep=1.5pt},
  box/.style={draw, rounded corners=1.5pt, fill=white, align=center,
    minimum width=1.9cm, minimum height=0.42cm, inner sep=1.5pt},
  frame/.style={draw, dashed, rounded corners=3pt, inner sep=4pt},
  ttl/.style={font=\scriptsize\bfseries, anchor=south west, inner sep=1pt},
  flow/.style={->, thick}]
\node[stp] (s1) at (0,0.3)    {\textbf{1} Verify\\projects};
\node[stp] (s2) at (2.05,0.3) {\textbf{2} Fetch issues\\(frozen snapshots)};
\node[stp] (s3) at (4.1,0.3)  {\textbf{3} Filter\\(staged funnel)};
\node[stp] (s4) at (6.15,0.3) {\textbf{4} Aggregate\\(densest quarter)};
\node[stp] (s5) at (0,-0.72)    {\textbf{5} Convert to\\\texttt{CaseInput}};
\node[stp] (s6) at (2.05,-0.72) {\textbf{6} Validate\\(runtime schema)};
\node[stp] (s7) at (4.1,-0.72)  {\textbf{7} Statistics\\(per-stage counts)};
\node[stp, fill=blue!6] (pool) at (6.15,-0.72) {Mined cases +\\issue pool};
\draw[flow] (s1) -- (s2);
\draw[flow] (s2) -- (s3);
\draw[flow] (s3) -- (s4);
\draw[flow] (s4.south) -- ++(0,-0.2) -| (s5.north);
\draw[flow] (s5) -- (s6);
\draw[flow] (s6) -- (s7);
\draw[flow] (s7) -- (pool);
\node[text=black!70, inner sep=1pt] at (3.07,-1.36)
  {fixed seed 42 \,$\cdot$\, offline replay (\texttt{-{}-skip-fetch}) \,$\cdot$\, funnel logged per stage};
\draw[dashed, rounded corners=3pt] (-0.98,-1.56) rectangle (7.13,0.98);
\node[ttl] at (-0.98,0.98) {(a) Mining ingestion pipeline (\S\ref{sec:mining})};
\node[box, minimum width=2.0cm, minimum height=1.4cm, fill=blue!6] (ci) at (8.9,-0.72)
  {\texttt{CaseInput}\\[1pt]\texttt{case\_name}\\\texttt{case\_description}\\\texttt{requirement}};
\node[box, minimum width=2.0cm, fill=gray!8] (hb) at (8.9,0.45) {5 hand-built cases};
\draw[flow] (hb) -- (ci);
\draw[flow] (pool.east) -- (ci.west);
\foreach \n/\t/\y in {a1/MARE/0.3, a2/iReDev/-0.18, a3/QUARE/-0.66}
  \node[box, minimum width=1.55cm] (\n) at (11.15,\y) {\t};
\node[box, minimum width=1.55cm, dashed, text=black!60] (a4) at (11.15,-1.14) {new adapter};
\node[frame, fit=(a1)(a4), inner sep=3pt] (fad) {};
\node[ttl] at (fad.north west) {Adapters};
\foreach \n/\t/\y in {p1/{P1 initial models}/0.3, p2/{P2 negotiation trace}/-0.18,
                      p3/{P3 integrated model}/-0.66, p4/{P4 verification}/-1.14}
  \node[box, minimum width=2.45cm] (\n) at (13.8,\y) {\t};
\node[frame, fit=(p1)(p4), inner sep=3pt] (fp) {};
\node[ttl] at (fp.north west) {Shared Phase 1--4 artifacts};
\node[box, minimum width=2.35cm, minimum height=0.9cm, fill=green!6] (met) at (16.85,0.02)
  {\textbf{Metrics}\\elements, MDC,\\Sem.\,F1, ISO ver., time};
\node[box, minimum width=2.35cm, minimum height=0.9cm, fill=green!6] (rr) at (16.85,-0.86)
  {\textbf{Run record}\\provenance, setting,\\model, seed, fallback};
\draw[flow] (ci.east |- 0,-0.42) -- (fad.west |- 0,-0.42);
\draw[flow] (fad.east |- 0,-0.42) -- (fp.west |- 0,-0.42);
\draw[flow] (fp.east |- met.west) -- (met.west);
\draw[flow] (fp.east |- rr.west) -- (rr.west);
\draw[dashed, rounded corners=3pt] (7.62,-1.56) rectangle (18.15,0.98);
\node[ttl] at (7.62,0.98) {(b) Run harness and shared contract (\S\ref{sec:overview})};
\end{tikzpicture}}
\caption{\tool{} architecture. (a)~Seven-step pipeline turning GitHub issues into
validated \texttt{CaseInput} cases. (b)~Any registered adapter consumes mined or
hand-built cases and emits shared Phase~1--4 artifacts, which the harness scores and records.}
\label{fig:arch}
\end{figure*}

\section{Introduction}
LLM-based multi-agent frameworks for requirements engineering (RE) are becoming
common. In these systems, role-specialized agents elicit, model, negotiate, and verify
requirements. Recent examples such as MARE~\cite{mare}, iReDev~\cite{iredev}, and
QUARE~\cite{quare} differ in agent roles, negotiation protocols, and produced
artifacts, which makes a basic question hard to answer: under identical conditions,
how do these frameworks actually compare?

Answering it is hard for two reasons. First, each framework reports results from
its own pipeline, cases, and metrics. MARE scores generated requirements models with
precision/recall against modeling baselines and rates its specifications by human
judgment~\cite{mare}; iReDev evaluates its artifacts on ten systems from a public
requirements-development dataset with its own metrics~\cite{iredev}. In our QUARE study~\cite{quare} we
reimplemented both from their papers to run them side by side. Second, the evaluation
inputs are small, curated sets of system descriptions, such as the ATM and library
systems drawn from prior literature~\cite{mare}; they are not mined from software
repositories with recorded provenance, and they do not grow as new projects appear.
Results on such inputs may not reflect how frameworks handle requirements as they
arise in active projects, which limits external validity; mining real issue
trackers offers a natural way to close this gap.

We address both obstacles with \tool, a reusable benchmarking infrastructure whose
contributions are:
\begin{enumerate}
  \item \textbf{A shared run harness and contract} that executes any conforming
  multi-agent RE framework behind one input\,/\,phase-artifact\,/\,run-record
  boundary and scores it with a common metric suite, so heterogeneous frameworks are
  run and compared under identical conditions (\S\ref{sec:overview}).
  \item \textbf{An adapter extension mechanism}: a new framework joins the comparison
  by implementing a single adapter against the contract, with no change to the cases,
  the metrics, or the other adapters (\S\ref{sec:overview}).
  \item \textbf{A reproducible GitHub issue-mining pipeline} that constructs
  benchmark cases from real feature-request issues with a fixed seed, frozen API
  snapshots, and per-stage funnel logging, so the corpus grows without hand-crafting
  (\S\ref{sec:mining}).
  \item \textbf{A versioned, compact case suite} (v1.0: five mined cases spanning
  three languages and five domains, plus five hand-built cases); the contribution is
  the mining-and-adapter extension mechanism, not raw corpus size (\S\ref{sec:dataset}).
  \item \textbf{Reference runs and reuse evidence}: measured runs of the three
  bundled frameworks demonstrating the harness end to end (\S\ref{sec:usage}), and
  reuse across our group's studies, including an adapter that reuses the
  originating pipeline while replacing its negotiation stage (see ``Relation to
  our prior work'' below).
\end{enumerate}

\tool{} is available for review at
\url{https://github.com/haowei614/OpenRE-Bench-Artifact/tree/msr2027-v1.0}, and
this exact release is permanently archived on Zenodo under
DOI~\href{https://doi.org/10.5281/zenodo.23875694}{10.5281/zenodo.23875694}.

\paragraph{Relation to our prior work}\label{sec:intro-reuse} The run harness and
metric suite began as the evaluation back-end of QUARE~\cite{quare}, our
multi-agent RE framework. This paper is the first to describe \tool{} as a
standalone, reusable tool, and it contributes components that were not part of
QUARE and are previously unpublished: the GitHub issue-mining pipeline, the five
mined cases and the released pool of \auditPoolN{} filtered issues, the filter audit, and
the reference runs on the mined cases. The infrastructure has since been reused
across our group: ConfigRE~\cite{configre} reused the harness for process-level
agent-configuration experiments, and ArgRE~\cite{argre} joined through the
adapter. ArgRE adapts QUARE's roles and Phases~1, 3, 4, and 5 and replaces Phase~2
(QUARE's Phase~5 only generates final documentation, emitted as optional artifacts
outside the scored Phase~1--4 contract); its evaluation compares it with MARE, iReDev, and an ablation of ArgRE itself. In
both cases the harness and the other adapters were used unchanged.

\paragraph{Related work} Three lines of work are related. (i)~\emph{LLM/agent benchmarks for
software engineering} such as SWE-bench~\cite{swebench} mine real GitHub issues but
target code patches, not requirements artifacts. (ii)~\emph{RE datasets} such as
PROMISE NFR~\cite{promisenfr} and PURE~\cite{pure} provide requirements text for
classification or extraction, but are static corpora, not an executable harness.
(iii)~\emph{Issue-mining infrastructure and tooling} in MSR, such as
GHTorrent~\cite{ghtorrent} (a continuous GitHub mirror and tool suite) and
Ticket~Tagger~\cite{issuetypes} (a GitHub app for issue-type labeling), support
repository mining or issue classification but are not executable framework
comparisons. Prior multi-agent RE frameworks are each evaluated on their own
runtime and mostly curated specifications (MARE~\cite{mare},
iReDev~\cite{iredev}; MARE also uses a PURE-derived public dataset), so their
numbers are not comparable. \tool{} differs by combining an executable,
framework-agnostic harness with a repository-mining pipeline that builds the
cases it runs.

\section{\tool{} Overview}\label{sec:overview}
\tool{} treats each RE framework as an executable adapter behind a shared runtime
boundary (Fig.~\ref{fig:arch}b): it standardizes the execution and evaluation interface (input, Phase~1--4
artifacts, metrics, provenance), not the frameworks' internal architectures or native
representations.

\paragraph{Shared input contract} Every adapter consumes the same \texttt{CaseInput}
(\texttt{case\_name}, \texttt{case\_description}, \texttt{requirement}). Cases are
discovered by convention (\texttt{*\_input.json}), so adding a hand-built or mined
case requires no code change.

\paragraph{Adapters} The three bundled adapters use different role sets and
negotiation policies: MARE has five task-specialized agents (Stakeholders,
Collector, Modeler, Checker, Documenter), iReDev six knowledge-driven agents
(Interviewer, EndUser, Deployer, Analyst, Archivist, Reviewer), and QUARE five
quality-axis agents (Safety, Efficiency, Green, Trustworthiness, Responsibility)
plus an Orchestrator for dialectical negotiation. All emit the same shared Phase~1--4
artifacts and a common run-record capturing provenance and comparability metadata.

\begin{table*}[t]
\centering
\caption{Mining funnel (issues surviving each stage; Raw: 600 most-recent
label-matching issues per project) and resulting cases (window, \#issues,
\texttt{\upshape requirement} words; cap 120/issue). Label and signal stages are no-ops here
and omitted: the fetch queries by label, and the signal gate is off in v1.0.}
\label{tab:funnel}
\footnotesize
\begin{tabular}{@{}lllrrrrrrlrr@{}}
\toprule
Project & Lang. & Domain & Raw & State & Length & Bug & Thread cap & Dedup & Window & \#Iss. & Words\\
\midrule
scikit-learn/scikit-learn & Python & ML Library        & 600 & 425 & 344 & 267 & 261 & 261 & 2021Q4 & 4 & 438\\
huggingface/transformers  & Python & ML / NLP          & 600 & 423 & 380 & 367 & 363 & 358 & 2024Q3 & 4 & 420\\
microsoft/vscode          & TS     & Editor / DevTool  & 600 & 399 & 304 & 258 & 258 & 258 & 2026Q3 & 3 & 372\\
elastic/elasticsearch     & Java   & Database / Search & 600 & 203 & 164 & 150 & 150 & 150 & 2026Q1 & 4 & 429\\
pandas-dev/pandas         & Python & Data Processing   & 600 & 446 & 442 & 370 & 370 & 369 & 2023Q3 & 3 & 368\\
\midrule
\textbf{Total} & & & \textbf{3{,}000} & \textbf{1{,}896} & \textbf{1{,}634} & \textbf{1{,}412} & \textbf{1{,}402} & \textbf{1{,}396} & 5 cases & \textbf{18} & \\
\bottomrule
\end{tabular}
\end{table*}

\paragraph{Shared representation and phases} Outputs use a KAOS-inspired~\cite{kaos}
shared encoding (Goal, Softgoal, Obstacle, Agent, plus Task and Resource from
$i^*$-style modeling), and all adapters emit Phase~1 (initial models), Phase~2
(negotiation/debate trace), Phase~3 (integrated model), and Phase~4
(verification), so heterogeneous outputs become comparable.

\paragraph{Settings ladder} A configuration ladder runs from a deterministic, key-free
\texttt{single\_agent} baseline up to \texttt{negotiation\_integration\_verification},
so reviewers run a cheap offline smoke check or a full LLM experiment from the same
entry point.

\paragraph{Metrics} Computed identically for every adapter: requirement and
integrated-model element counts (a descriptive size of the shared KAOS encoding);
MDC, the mean distance of Phase~3 elements to their centroid in a five-dimensional
quality space (a dispersion measure, where higher means more spread-out
elaboration); BERTScore~\cite{bertscore}~F1 of Phase~3 versus that run's own
Phase~1 as within-run semantic preservation; an
ISO/IEC/IEEE~29148 verifiability score on a $1$--$5$ scale, computed
deterministically from structural-compliance signals (phase-level deterministic
validity and compliance coverage) rather than by an LLM judge, so it is
reproducible from the artifacts; and wall-clock time. A two-annotator study of
verifiability ships in the artifact (quadratic-weighted $\kappa=0.76$).

\paragraph{Evaluation protocol} Given a \texttt{CaseInput}, a framework emits
Phase~1--4 artifacts that the harness scores under a fixed setting, model,
temperature, and seed. Table~\ref{tab:results} (\S\ref{sec:usage}) reports
reference runs under this protocol; they demonstrate the tool and do not rank the
frameworks (\S\ref{sec:limits}).

\paragraph{Versioning and reproducibility} Each case suite is a frozen, citable,
DOI-versioned snapshot (v1.0: five mined plus five hand-built cases), so cross-paper
results refer to the same version while the pipeline extends future ones. The
environment pins Python~3.13 with a \texttt{uv} lockfile and a Docker image, stores
per-run provenance, and is guarded by a 651-test suite. Adding a framework (\S\ref{sec:usage})
leaves cases, metrics, and other adapters untouched.

\section{Mining Ingestion Pipeline}\label{sec:mining}
The ingestion module converts real GitHub issues into
harness cases. It is organized as seven independently runnable, configurable steps
(Fig.~\ref{fig:arch}a)
(project verification, issue fetch, filtering, aggregation, schema conversion, case
validation, and statistics) so that each step's effect is auditable and reproducible.
The filtering step itself applies a staged funnel, whose per-stage survivor
counts are reported in Table~\ref{tab:funnel} (\S\ref{sec:dataset}).

\paragraph{Project selection} Candidate projects must pass explicit thresholds on
stars, recent activity, and open/closed issue volume, and are chosen for
language/domain diversity. A verification step queries each project's issue endpoint
and rejects unminable projects (e.g.\ one routing issues to an external tracker,
with GitHub issues disabled) rather than silently yielding an empty corpus. The final
five projects (\texttt{scikit-learn}, \texttt{transformers}, \texttt{vscode},
\texttt{elasticsearch}, \texttt{pandas}) span three languages (Python, TypeScript,
Java) and five domains.

\paragraph{Issue filtering} Each project's issues pass through a staged funnel:
(1) label allow-list, (2) state/reactions gate, (3) minimum body length, (4) bug
exclusion, (5)~optional requirement-signal keyword gate (default off in v1.0;
see audit below), (6) a mega-thread comment cap, and (7) TF-IDF cosine
deduplication. A reproducibility subtlety we document is that the GitHub
\texttt{labels} query parameter uses conjunctive semantics; to obtain the
intended union we query each label separately and merge by issue number.

\paragraph{Aggregation} Surviving issues are clustered by calendar-quarter time
window; the pipeline selects the densest window and the top issues by community
reactions, applying a per-issue word cap so that a mined case matches the textual
scale of the hand-built cases.

\paragraph{Conversion and validation} Each aggregated case is emitted as a
\texttt{CaseInput} JSON and validated with the same schema model the harness loads at
runtime. Any case that validates is therefore runnable; source issue URLs are retained
in a metadata sidecar for traceability.

\paragraph{Reproducibility guarantees} All sampling and clustering use a fixed seed
(42); raw API responses are committed as timestamped snapshots, so the pipeline
re-derives the same cases offline (\texttt{run\_all.py -{}-skip-fetch}); the API token
is read from the environment or a git-ignored file, never committed; every stage logs
input/output counts.

\begin{table*}[t]
\centering
\caption{Reference runs on the mined cases (\texttt{\upshape claude-haiku-4-5},
\texttt{\upshape negotiation\_integration\_verification}): mean $\pm$ SD, 5 cases $\times$ 3 seeds
($n{=}15$). Metrics as in \S\,II-e.}
\label{tab:results}
\footnotesize
\begin{tabular}{@{}lrrrrr@{}}
\toprule
Adapter & P3 elems & MDC & Sem.\,F1 & ISO ver. & Time (s) \\
\midrule
MARE   & 24.9\,$\pm$\,0.3 & 0.793\,$\pm$\,0.052 & 0.924\,$\pm$\,0.011 & 3.02\,$\pm$\,0.04 & 37.2\,$\pm$\,2.8 \\
iReDev & 30.0\,$\pm$\,0.0 & 0.743\,$\pm$\,0.036 & 0.943\,$\pm$\,0.004 & 3.03\,$\pm$\,0.07 & 248.5\,$\pm$\,14.4 \\
QUARE  & 35.0\,$\pm$\,0.0 & 0.675\,$\pm$\,0.002 & 0.951\,$\pm$\,0.003 & 4.82\,$\pm$\,0.17 & 109.7\,$\pm$\,6.0 \\
\bottomrule
\end{tabular}
\end{table*}

\section{Mined Dataset and Statistics}\label{sec:dataset}
Table~\ref{tab:funnel} reports the full mining funnel: from 3{,}000 fetched issues the
pipeline retains 18 across five cases, with the state/reactions gate removing
the most.

The last three columns of Table~\ref{tab:funnel} describe the resulting cases. Mined cases (368--438 words) are modestly
larger than the five hand-built cases (58--337 words) but of the same order,
and load under the identical \texttt{CaseInput} schema. The reduction from
\auditPoolN{} survivors to 18 issues is aggregation, not further filtering (the
densest-window selection above). We release the full \auditPoolN{}-issue pool as an
issue-level dataset (repository, number, URL, title, labels, timestamps,
reactions, per-stage markers, and a flag for the 18 aggregated issues).

Because the 18 aggregated issues were selected by community reactions, they cannot by
themselves validate the filter. We therefore audited an earlier configuration that
enabled the requirement-signal keyword gate: a seeded sampler (seed~42) drew 50 issues
from the \auditPoolNOne{} survivors of that gate and 30 from issues the gate dropped,
with a one-page annotation guideline and two independent raters
($\kappa=\auditKappa{}$, \auditDisagree{} disagreements, later majority-adjudicated).
On the adjudicated retained sample, precision is \auditPrec{} (Wilson 95\%\ CI
\auditPrecCI; $n=\auditRetainedN$). On the adjudicated dropped sample we report the
false-omission rate: the share of dropped issues that are still genuine requirements
(i.e., $\mathrm{FN}/(\mathrm{FN}+\mathrm{TN})$; not the miss rate
$\mathrm{FN}/(\mathrm{FN}+\mathrm{TP})$), which is \auditFOR{}
(\auditFORCI; $n=\auditRejectedN$). Consensus-only rates (disagreements excluded) are
\auditPrecConsensus{}/\auditFORConsensus{}. The gate therefore kept a moderately pure
set but discarded mostly real requirements, so it lacked discriminative power. v1.0
consequently disables the gate by default while retaining it as a config flag for
ablation; the released pool has size
$\auditPoolN{}=\auditPoolNOne{}+\auditPoolNTwo{}$. Reusing the same labels as a
two-stratum estimate of pool precision
($N_1=\auditPoolNOne{}$ signal-matching, $N_2=\auditPoolNTwo{}$ signal-absent;
$P=(N_1/N)p_1+(N_2/N)p_2$) yields \auditPoolPrec{} (normal-approx 95\%\ CI
\auditPoolPrecCI). All audit statistics are recomputable via \texttt{scripts/audit\_stats.py}.

\section{Demonstration and Reference Runs}\label{sec:usage}

\paragraph{Using and extending \tool} The core workflows are one command each and
run in seconds with no API key or cost.
\begin{lstlisting}
uv sync --all-groups   # pinned Python 3.13 env
# (a) Run a case offline, no key (smoke test):
uv run openre_bench --run-case --system quare \
  --case-input data/case_studies/vscode_2026q3_input.json \
  --artifacts-dir out --setting single_agent
# (b) Replay mining offline (seed 42):
uv run python ingestion/run_all.py --skip-fetch
# (c) Recompute the reference table from runs:
uv run python scripts/recompute_reference_table.py
\end{lstlisting}
A new framework joins by implementing one adapter module that exposes three
callables against the shared contract:
\texttt{build\_phase1} (Phase~1 elements, runtime semantics, and run metadata),
\texttt{build\_negotiation\_history} (the Phase~2 trace), and
\texttt{build\_optional\_artifacts}, plus one registry line. No case, metric, or
other adapter changes are required.

\paragraph{Reference runs} To demonstrate the harness end to end, we ran every mined
case through all three adapters with a real LLM (\texttt{claude-haiku-4-5},
temperature $0.7$, round cap $3$, seeds $101/202/303$, four settings). The harness
flags any run that falls back to the deterministic scaffold as
\emph{fallback-tainted}; all reported runs are LLM-backed and pass the comparison
validator. Table~\ref{tab:results} reports the
negotiation-integration-verification setting as mean\,$\pm$\,SD over five cases and
three seeds.
For \texttt{sklearn\_2021q4} and \texttt{pandas\_2023q3} the runs use the v1.0 (signal-off) case
texts; for the other three cases the v1.0 texts are byte-identical to those of
the earlier runs, which are reused unchanged; all values are recomputable (step~(c)).
The runs show that heterogeneous adapters execute under one protocol, emit
contract-valid artifacts, and yield seed-stable, recomputable metric profiles.
Element counts are largely fixed by each
adapter's deterministic Phase~1 KAOS scaffold (MARE~25, iReDev~30, QUARE~35 slots;
observed P3 means $24.9\pm0.3$ / $30.0\pm0.0$ / $35.0\pm0.0$), so the LLM mainly fills
slot content (semantic preservation and structural verifiability) rather than slot
count. On the five hand-built cases the same protocol reproduces the same metric
profile (\texttt{experiment\_outputs/handbuilt-claude/}), which suggests the pattern
is not tied only to the mined corpus.

\paragraph{Cost} Reproduction is inexpensive: the single-seed matrix (60 runs, 752
LLM calls) costs \$3.59 on \texttt{claude-haiku-4-5} and the full three-seed matrix
(180 runs, 2{,}266 calls) \$10.96 (from the packaged \texttt{usage\_report.json}); the
key-free setting reproduces the pipeline offline at zero cost.

\section{Limitations and Threats to Validity}\label{sec:limits}
\emph{Adapter-induced and representation-induced bias.}
\tool{} keeps the evaluation interface consistent, but does not eliminate all
representation-induced biases. A framework is comparable only if its output maps to
the shared Phase~1--4~/~KAOS contract; that mapping can mask native behavior, and the
bundled adapters' deterministic Phase~1 scaffolds fix element counts before the LLM
runs. Content-facing scores (semantic preservation vs.\ a run's own Phase~1;
structural ISO verifiability) remain useful as protocol diagnostics, yet must not be
read as scaffold-independent rankings. MARE and iReDev are reimplemented from
their papers, so absolute fidelity to the originals is
a remaining threat. \emph{Training-data contamination.} For
\texttt{claude-haiku-4-5} (training cutoff July~2025), three mined windows predate the
cutoff and two postdate it; re-mining a post-cutoff window yields a refreshable,
contamination-free suite on demand. \emph{Aggregation and sampling.} Calendar-quarter
grouping can mix unrelated topics; the fetch is capped at 600 recent label-matching
issues per project; the signal gate is off by default after the \S\ref{sec:dataset}
audit; fixed seeds, model identifiers, and temperatures aid comparability, but exact
LLM outputs are not guaranteed to reproduce.
\emph{Suite size.} v1.0 is ten cases; the frozen release is kept separate from the
extensible pipeline.

\paragraph{Future extensions} Planned extensions include other issue trackers (Jira,
GitLab) behind the same fetch/filter interface; topic-cluster aggregation as an
alternative to time windows; more framework adapters via the one-module contract; and
periodic post-cutoff releases to keep a contamination-free suite current.

\section{Availability and Conclusion}\label{sec:concl}
\tool{} releases its code under AGPL-3.0 and its curated data under CC~BY~4.0, with a
pinned environment, Docker image, frozen snapshots, and citation metadata. Embedded
GitHub issue text stays its authors' copyright, redistributed for research use with
source URLs retained and \texttt{@}-mentions de-identified. New frameworks can be
added as adapters and new cases mined from further projects, so the suite can grow
with community contributions.


\end{document}